\documentclass[smallextended]{svjour3}
\usepackage{amsfonts}
\usepackage{bm}
\usepackage{verbatim}
\usepackage{color}
\usepackage{graphicx}

\newcommand{\im}{i}
\newcommand{\unitvec}[1]{\hat{\mathbf{#1}}}

\def\onlinecite#1{\cite{#1}}

\begin{document}

\title{Spin, orbital, Weyl and other glasses in topological superfluids}

\author{G.E.~Volovik \and J.~Rysti \and J.T.~M\"akinen \and V.B.~Eltsov}

\institute{G.E.~Volovik, J.~Rysti, J.T.~M\"akinen, V.B.~Eltsov \at
Department of Applied Physics, Aalto University, P.O. Box 15100, FI-00076
Aalto, Finland \and
G.E.~Volovik \at
Landau Institute for Theoretical Physics, acad. Semyonov av., 1a,
142432, Chernogolovka, Russia}

\date{Received: \today / Accepted: date}

\maketitle

\begin{abstract}
One of the most spectacular discoveries made in superfluid $^3$He
confined in a nanostructured material like aerogel or nafen was the observation of the destruction of the long-range orientational order by a weak random  anisotropy. The quenched random
anisotropy provided by the confining material strands produces several different
glass states resolved in NMR experiments  in the chiral superfluid
$^3$He-A and in the time-reversal-invariant polar phase. The smooth
textures of spin and orbital order parameters in these glasses can be
characterized in terms of the randomly  distributed  topological charges,
which describe skyrmions, spin vortices and hopfions. In addition, in
these skyrmion glasses the momentum-space topological invariants are
randomly distributed in space. The Chern mosaic, Weyl glass, torsion
glass and other exotic topological sates are examples of close
connections between the real-space and momentum-space topologies in
superfluid $^3$He phases in aerogel.
\end{abstract}


\section{Introduction}

 The spin triplet $p$-wave superfluid phases of liquid $^3$He \cite{VollhardtWolfle1990} immersed in the aerogel matrix provide the arena for experimental and theoretical investigations of  different types of spin and orbital orientational  disorder, induced by the quenched orientational disorder of the aerogel strands. 
 Especially interesting phenomena are realized in the chiral superfluid $^3$He-A phase, which in addition to superfluidity has the signatures of the spin nematic \cite{AndreevMarchenko1980,Andreev1984} and orbital ferromagnet.
 One of the most spectacular discoveries was the observation of the destruction of the long-range orientational order in $^3$He-A by a weak random  anisotropy \cite{Dmitriev2010} -- the so-called  Larkin-Imry-Ma (LIM) effect \cite{Larkin1970,ImryMa,Volovik1996} (see also review paper \cite{Halperin2018}).
 This is the orbital glass state of the chiral superfluid $^3$He-A -- the bulk 3D topological system with smooth disorder in the field of the orbital vector $\hat{\bf l}$, which describes the orientation of the orbital magnetization of Cooper pairs in this chiral liquid. The smooth  texture of the $\hat{\bf l}$-vector can be characterized by the  integer valued topological charges valid for the soft topological objects, such as 2D and 3D skyrmions \cite{Skyrme1958}, merons, continuous vortices \cite{Salomaa1987}, topological solitons, domain walls, monopoles and hedgehogs. Following the notations of Refs.~\cite{Chudnovsky2017,Chudnovsky2018}, the LIM orbital glass state can be called the {\it intrinsic orbital skyrmion glass}.

The intrinsic orbital glass state persists even when the global anisotropy of the aerogel strands is present. In the polar-distorted $^3$He-A phase (PdA phase), which is formed  in the  aerogel with "nematically ordered" strands, provided by the commercially-available nafen material \cite{nafen}, the 2D LIM state is observed with the disordered planar texture of the $\hat{\bf l}$-vector \cite{Askhadullin2014}.

The intrinsic orbital glass is realized as an equilibrium state. Whether it is the true glass state or the orbital liquid is an open question. The $^3$He-A in aerogel may have many degenerate ground states (or nearly degenerate states)  with the rare events of the transitions between the states.  All these states have smaller energy compared to the ordered state of the orbital ferromagnet, and thus are not able to relax to the ordered state with long-range order.  

In addition, there are the topological excited states on the background of the intrinsic LIM glass. 
In particular,  it was found that the aerogel strands strongly pin the singular topological defects (with hard cores), such as quantized vortices and  half-quantum vortices -- Alice strings \cite{Autti2016a}.
The disordered state which contains pinned vortices is obtained by the Kibble-Zurek  mechanism: vortex nucleation by fast cooling through $T_c$ \cite{Kibble1976,Zurek1985,Ruutu1996}. 

\begin{figure}[t]
\includegraphics[width=\columnwidth]{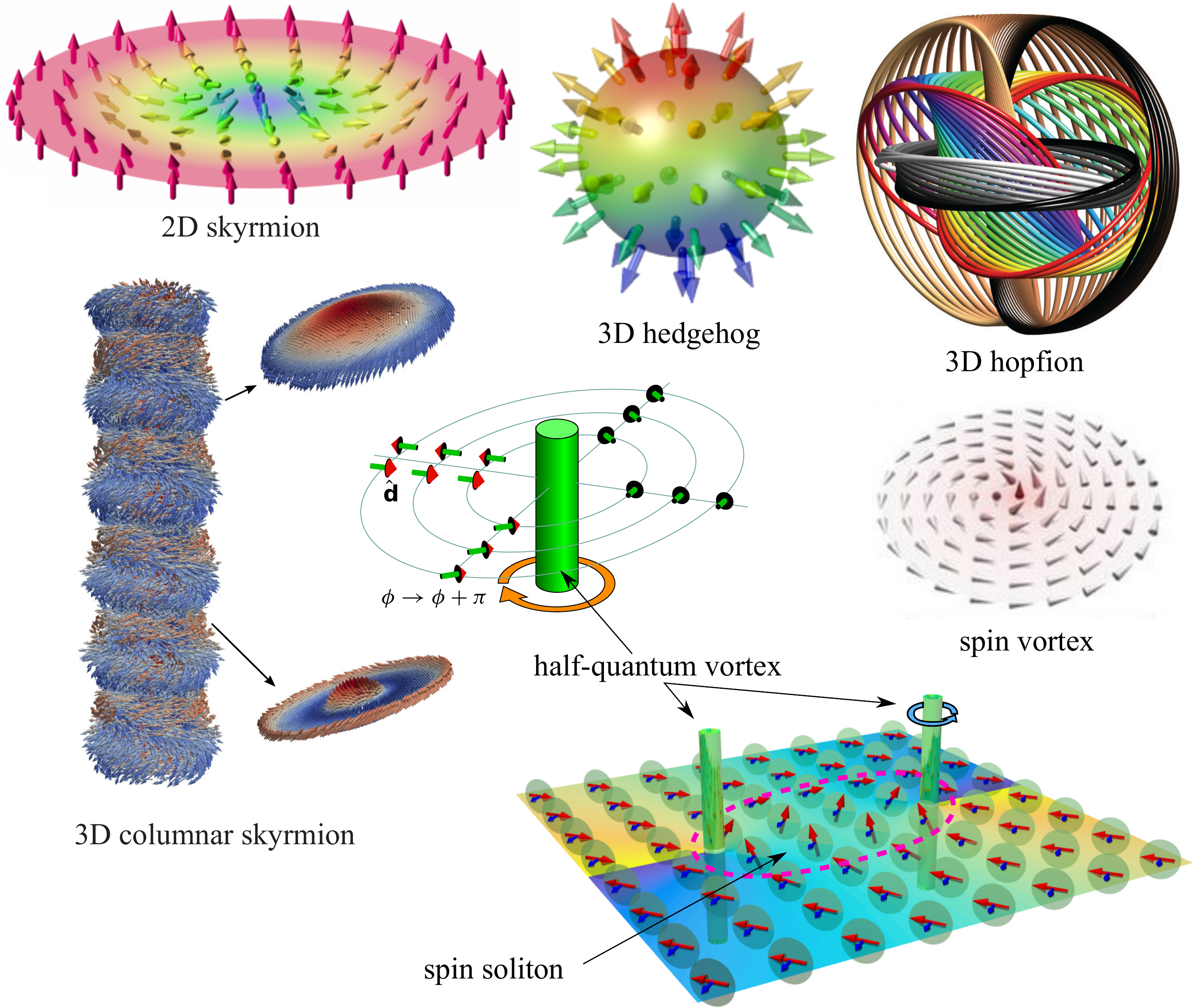}
\caption{(color online) Zoo of topological defects pinned by aerogel.
Hard-core topological defects (half-quantum vortices and hedgehogs), intermediate-core defects (spin vortices), and soft objects (skyrmions, hopfions and spin solitons).
Images adapted: skyrmion and hedgehog from Ref.~\onlinecite{skyrmion-im}, hopfion from Ref.~\onlinecite{hopfion-im}, spin vortex from Ref.~\onlinecite{spinvortex-im} and columnar skyrmion from Ref.~\onlinecite{colskyrm-im}.
}
\label{objects}
\end{figure}

The formed excited state can be called the {\it vortex glass}. This vortex glass is very different from the Larkin vortex glass in superconductors, where vortices have preferred  orientation of magnetic fluxes along the magnetic field.
In the isotropic aerogel vortices have random  orientation of vortex lines. In the aerogel with preferable orientations of the strands the vortices form the disordered Ising glass with the random distributions of the 
winding numbers $N=+1$ and $N=-1$, or $N=+1/2$ and $N=-1/2$ in case of half-quantum vortices in Fig.~\ref{objects}.

\begin{figure}[t]
\includegraphics[width=\columnwidth]{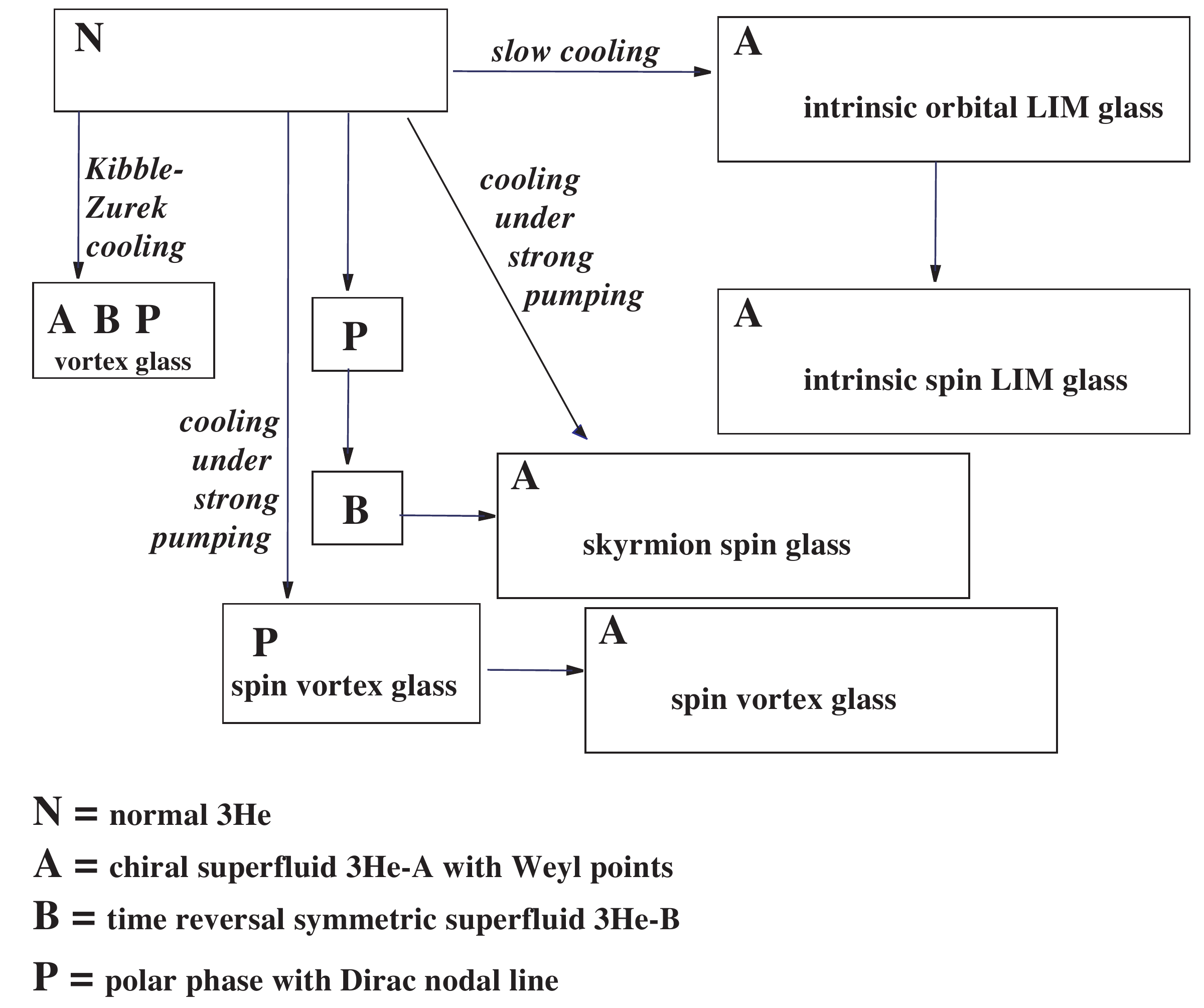}
\caption{Variety of glass states obtained by different protocols.
}
\label{Table}
\end{figure}

The aerogel and nafen also pin the topological defects with the cores of intermediate sizes, such as spin vortices  in Fig.~\ref{objects}. All these possibilities, in addition to the fully equilibrium LIM state, give rise to a zoo of quasiequilibrium glass states with different types of the pinned topological excitations. These states can be obtained using different protocols, see Fig.~\ref{Table}.  

The rich glass states in superfluid phases of $^3$He could be useful for studies of different problems related to spin glasses \cite{Villain1977,DzyaloshinskiiVolovik1978,Binder1986}.
Our measurements in superfluid $^3$He confined within nafen, Fig.~\ref{fig:exper}, demonstrate that there are at least three types of spin-glass states with different NMR signatures. One of them is the equilibrium spin-glass state: due to spin-orbit interaction the orbital glass serves as the quenched orientational disorder acting on the spin-nematic vector $\hat{\bf d}$.  The $\unitvec{d}$ vector is a unit vector (director) along the spontaneous uniaxial spin anisotropy of the A-phase. As a result  the LIM state of the $\hat{\bf d}$ vector is formed with the characteristic LIM scale larger than LIM scale in the orbital glass.

Other possible spin-glass states can be considered as the topological excited states of the spin LIM.
These are the spin-skyrmion and spin-vortex glass states, which pin the spin skyrmions
and spin vortices in Fig.~\ref{objects}.
The spin-skyrmion glass is formed when the transition from the normal liquid to the A-phase is accompanied by strong magnetic perturbations \cite{Dmitriev2010}. The same state is obtained by the first-order transition from the B-phase to the A-phase. As distinct from the orbital glass, which is realized as the equilibrium state (or as a manifold of nearly degenerate states), the spin glass in $^3$He-A is not an equilibrium state and can be annealed. Following notations \cite{Chudnovsky2017,Chudnovsky2018}, we call it {\it spin-skyrmion glass}.

The {\it spin-vortex glass} states are obtained by cooling  from the normal liquid to the polar phase under strong magnetic perturbations. This spin glass can be represented as a chaotic system of spin vortices pinned by the aerogel. 

Here we consider how different topological charges characterize different types of glass states, and apply the simplest Larkin-Imry-Ma arguments to describe the properties of these glasses leaving a more detailed consideration for the future.

Topology of skyrmion glasses are discussed  in Secs. \ref{2DSkyrmGlassSec}
and \ref{SkyrmGlassSec}; the orbital LIM glass is in Sec.~\ref{OrbGlassSec}; the spin glasses including spin-vortex glass  are in Sec.~\ref{SpinGlassSec};
the vortex glasses are in Sec. \ref{VortexGlassSec}: the combined effect of real and momentum space topology is in Sec.~\ref{FermGlassSec}.

\section{2D skyrmion glass}
\label{2DSkyrmGlassSec}

\subsection{Skyrmionic topology of 2D glass}

Both the equilibrium LIM glasses emerging in magnets with random anisotropy or with random field and the quasiequilibrium skyrmionic glasses can be described in terms  of the skyrmionic topological invariants. For the 2D  Heisenberg  spin glasses the relevant invariant is $\pi_2(S^2)=Z$, see e.g. Refs. \onlinecite{Chudnovsky2017,Chudnovsky2018}:
\begin{equation}
Q_2=\int d{\bf S}\cdot {\bf q}_2(x) \,\,,\,\,  q_2^i=
\frac{e^{ikl}}{8\pi}\,\hat{\bf s}\cdot (\partial_k \hat{\bf s}\times 
\partial_l \hat{\bf s})
\,.
\label{TopInv2}
\end{equation}
Here $\hat{\bf s}$ is the unit vector of magnetization in ferromagnets. It should be substituted by the nematic vector (director) $\hat{\bf d}$ in the chiral A phase and in the polar phase of superfluid $^3$He, by the orbital vector $\hat{\bf l}$ in the A phase, and by the antiferromagnetic vector in antiferromagnets.

\subsection{Fluctuations of topological charges in 2D skyrmion glass}

Consider the mean square fluctuations of the total topological charge,  $\langle Q_2^2\rangle$, assuming that  
 $\langle Q_2\rangle=0$. Of course, if one fixes the topological charge by boundary conditions, then
$Q_2$ does not fluctuate. If the order parameter is fixed at
the boundaries,
then the change of the total topological charge of the whole texture is
possible only by singular process of creation of topological charge inside
the sample -- this is the instanton process, in which the system crosses the singularity in the $2+1$ spacetime -- the spacetime hedgehog with topological charge $\pi_2(S^2)=Z$.
If the boundary conditions are fixed and the instanton processes are ignored,  one can choose the  finite region of the intermediate size $L$ inside the glass, which is much smaller than the dimension of the system and much larger than the LIM scale $\xi_{\rm LIM}$. Then the total charge $Q_2$ in this region is fluctuating. In general one may expect 
\begin{equation}
\langle Q_2^2\rangle \sim (L/\xi_{\rm LIM} )^m
\,.
\label{ScaleQ2}
\end{equation}
In the simplest model of the Gaussian distribution, the
$\langle Q_2^2\rangle$ is proportional to the area $S$ of the region, and one has $m=2$:
 \begin{eqnarray}
 \langle Q_2^2\rangle \sim S \langle q_2^2\rangle\xi_{\rm LIM}^2 \sim (L/\xi_{\rm LIM} )^2
\,,
\label{Q_2^2}
\end{eqnarray}
where we use the dimensional analysis result
\begin{equation}
\langle q_2^2\rangle\sim \xi_{\rm LIM}^{-4}
\,.
\label{q^2}
\end{equation}

\section{3D skyrmion glasses}
\label{SkyrmGlassSec}

\subsection{Topology of 3D skyrmion glasses}

In the 3D Heisenberg  magnetic glasses and in the superfluid A and polar phases of $^3$He, the skyrmionic charge is described by $\pi_3(S^2)=Z$ topology. The topological charge is the Hopf invariant, which can be expressed  in terms of helicity of the effective gauge field \cite{VolovikMineev1977}:
\begin{equation}
Q_3=\int d^3x  q_3(x) \,\,,\,\,  q_3(x)=
\frac{e^{ikl}}{32\pi^2}\,A_i F_{kl}
\,,
\label{TopInv3}
\end{equation}
where the synthetic gauge field has the following connection to the vector field:
\begin{equation}
F_{kl}=\partial_k A_l- \partial_l A_k= \hat{\bf s}\cdot (\partial_k \hat{\bf s}\times 
\partial_l \hat{\bf s})
\label{GaugeField}
\end{equation}
For 3D ferromagnets, the synthetic gauge field ${\bf A}$ is the Berry phase field \cite{Volovik1987}.
In these solid-state magnetic materials, the particle-like topological excitations described by the Hopf invariant -- hopfions in Fig. \ref{objects} (or knots \cite{Tiurev2018}) -- are suggested for spintronics applications \cite{Smalyukh2018,Lake2018}.

\subsection{Fluctuations of Hopf topological charge in 3D skyrmion glass}

Let us consider fluctuations in the 3D skyrmion glasses, assuming that there are no singular defects (hedgehogs, strings and domain walls). The absence of the singular defects allows us to use for their description both topological charges, which characterize the continuous configurations, $Q_2$ and  the Hopf invariant, $Q_3= \int d^3x\, q_3(x)
\sim  \int d^3x\, {\bf A}\cdot(\nabla\times {\bf A})$. As before we consider fluctuations in the volume $V$ which is much smaller than the total volume of the sample and much larger than the volume of  LIM scale. If the regions with positive and negative $q_3(x)$ are
randomly distributed, and we assume the Gaussian distribution, then  $\langle Q_3^2\rangle$ is proportional to the volume $V$:
\begin{eqnarray}
\langle Q_3^2\rangle\sim V/\xi_{\rm LIM}^3
\,.
\label{Q_3^2a}
\end{eqnarray} 
On the other hand, as follows from Eq.~(\ref{TopInv3}), $\langle Q_3^2\rangle$ can be expressed in terms of the distributions of the synthetic gauge field ${\bf A}$ and the topological density $q_2$:
\begin{eqnarray}
\langle Q_3^2\rangle\sim V \xi_{\rm LIM}^3 \langle A^2\rangle\langle q_2^2\rangle 
\,.
\label{Q_3^2b}
\end{eqnarray}
This allows us to estimate the fluctuations of the effective gauge field, $\langle A^2\rangle$, using Eqs.(\ref{q^2}), (\ref{Q_3^2a}) and (\ref{Q_3^2b}): 
\begin{equation}
\langle A^2\rangle\sim \xi_{\rm LIM}^{-2}
\,.
\label{A^2}
\end{equation}

\subsection{Fluctuations of $Q_2$ topological charge in 3D skyrmion glass}

In the 3D skyrmion glass it is instructive to consider the fluctuations of the topological charge  $Q_2$. 
In 3D systems this invariant describes the columnar textures, the line object described by the mapping of the cross-section of the skyrmion to the sphere of the unit vector   $\hat{\bf l}$, which form the homotopy group   $\pi_2(S^2)=Z$, see columnar skyrmions in Fig. \ref{objects}.  For example, in $^3$He-A this texture represents the vortex line with the continuous order parameter \cite{Salomaa1987}. The similar linear skyrmions are formed in magnets with Dzyaloshinsky-Moriya interaction.
Let us consider fluctuations  $\langle Q_2^2\rangle$ in the 2D  cross-section of the 3D system. We again assume that the equilibrium LIM state is smooth and does not contain the singular structures, such as singular vortices and  hedgehogs. Then the integral (\ref{TopInv2}) over any closed surface, which is equal to the total topological charge of the hedgehogs inside the surface, is zero. This gives reduction of  $\langle Q_2^2\rangle$ imposed by the constraint: all the surfaces which have the common boundary have the same value of the topological invariant. The reduced scaling law for $\langle Q_2^2\rangle$ can be obtained from Eq.~(\ref{A^2}):
\begin{equation}
 \langle Q_2^2\rangle\sim \left(\oint {\bf A }\cdot d{\bf x}\right)^2 \sim \langle A^2\rangle L\xi_{\rm LIM} \sim L/\xi_{\rm LIM}
\,.
\label{q_23D}
\end{equation}
So in the pure 2D system one has Eq.~(\ref{Q_2^2}), which corresponds to $m=2$ in Eq.~(\ref{ScaleQ2}), while for the 3D systems 
one has Eq.~(\ref{q_23D}), which corresponds to $m=1$ in Eq.~(\ref{ScaleQ2}). The crossover from 2D to 3D takes place at 
the thickness $L_z \sim \xi_{\rm LIM}$.

\begin{figure}
\centerline{\includegraphics[width=\linewidth]{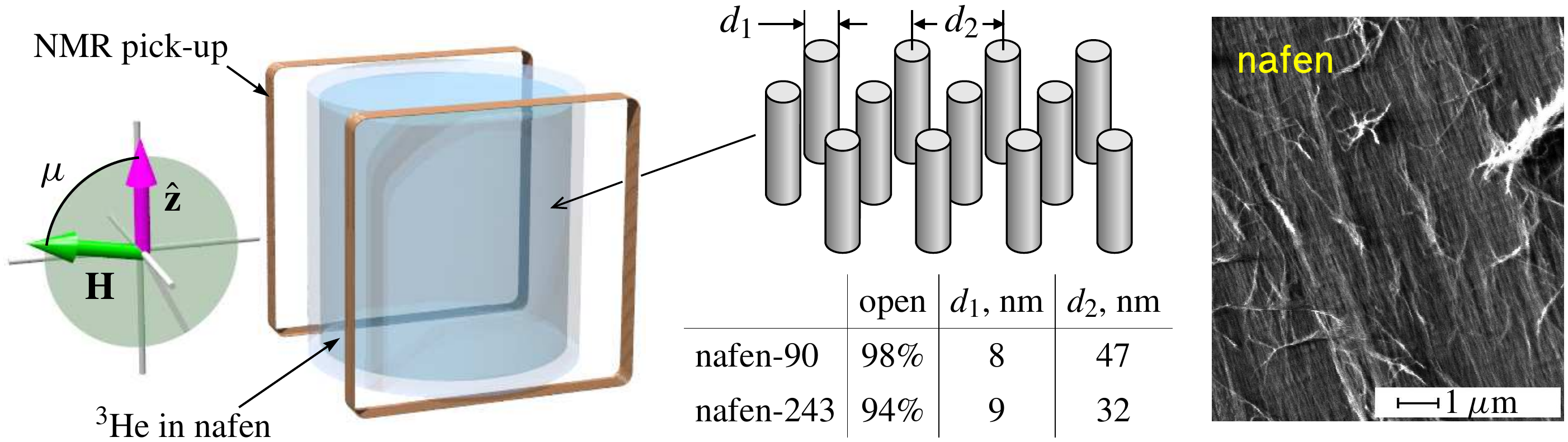}}
\caption{Sketch of the experimental setup for measurements of glass states of topological superfluid $^3$He confined in nafen nanostructured material.
\textit{(Right)} Nafen consists of nearly parallel Al$_2$O$_3$ strands, while liquid $^3$He penetrates between strands. Two different nafen variants with densities of 90 and 243\,mg/cm$^3$ have been used. The confinement parameters are shown in the middle \cite{nafen}. The nafen possesses two types of disorder which allows formation of glass states in topological superfluid: (i) orientational disorder in strands direction and (ii) density fluctuations (variation of $d_2$). \textit{(Left)} The nafen is placed into the cylindrical sample container (height 4\,mm and diameter 4\,mm) with the average strand direction along the cylinder's axis, denoted as $\hat{\mathbf{z}}$. The sample is surrounded by NMR pick-up coils and the static magnetic field can be oriented at an arbitrary angle $\mu$ with respect to nafen strands. To avoid formation of paramagnetic solid $^3$He on all surfaces, the sample is preplated by 2.5 atomic layers of $^4$He \cite{he4nafen}.}
\label{fig:exper}
\end{figure}

\section{Glasses in chiral superfluids}
\label{ChiralSuperfluidSec}

The spin-triplet $p$-wave order parameter $\Delta_{\alpha\beta}(\mathbf{k}) = (\im \sigma^2 \sigma^{\mu})_{\alpha\beta}k^iA_{\mu i}$ of chiral superfluid $^3$He-A is given by
\begin{equation}
A_{\mu i} =\Delta_{\rm A} e^{i\Phi} \hat{d}_{\mu}(\hat{\bf e}_1 + i \hat{\bf e}_2)_i
\,.
 \label{eq:OPHeA}
\end{equation}
Here $\hat{\bf d}$ is the the  spin-nematic vector; the unit vectors $\hat{\bf e}_1\perp \hat{\bf e}_2$ describe the orbital degrees of freedom.\footnote{Traditionally in $^3$He literature, the orbital vector $\unitvec{e}_1$ is named $\unitvec{m}$ and vector $\unitvec{e}_2$ is named $\unitvec{n}$. We use $\unitvec{e}_{1,2}$ notation to stress connection to tetrad field. Also the vector $\unitvec{e}$ introduced later in the polar phase is usually named $\unitvec{m}$.}  The unit vector  $\hat{\bf l}=\hat{\bf e}_1\times \hat{\bf e}_2$ plays several  roles in chiral superfluid: it shows the direction of the orbital angular momentum of Cooper pairs and thus determines the orbital magnetization of chiral superfluid; it determines the easy axis of the orbital anisotropy; it shows the direction to the Weyl nodes in the fermionic quasiparticle spectrum in momentum space; together with vectors $\hat{\bf e}_1$ and $\hat{\bf e}_2$ it forms the analog of the tetrad fields in general relativity; it is also responsible for continuous vorticity of the superflow velocity.

\begin{figure}
\centerline{\includegraphics[width=\linewidth]{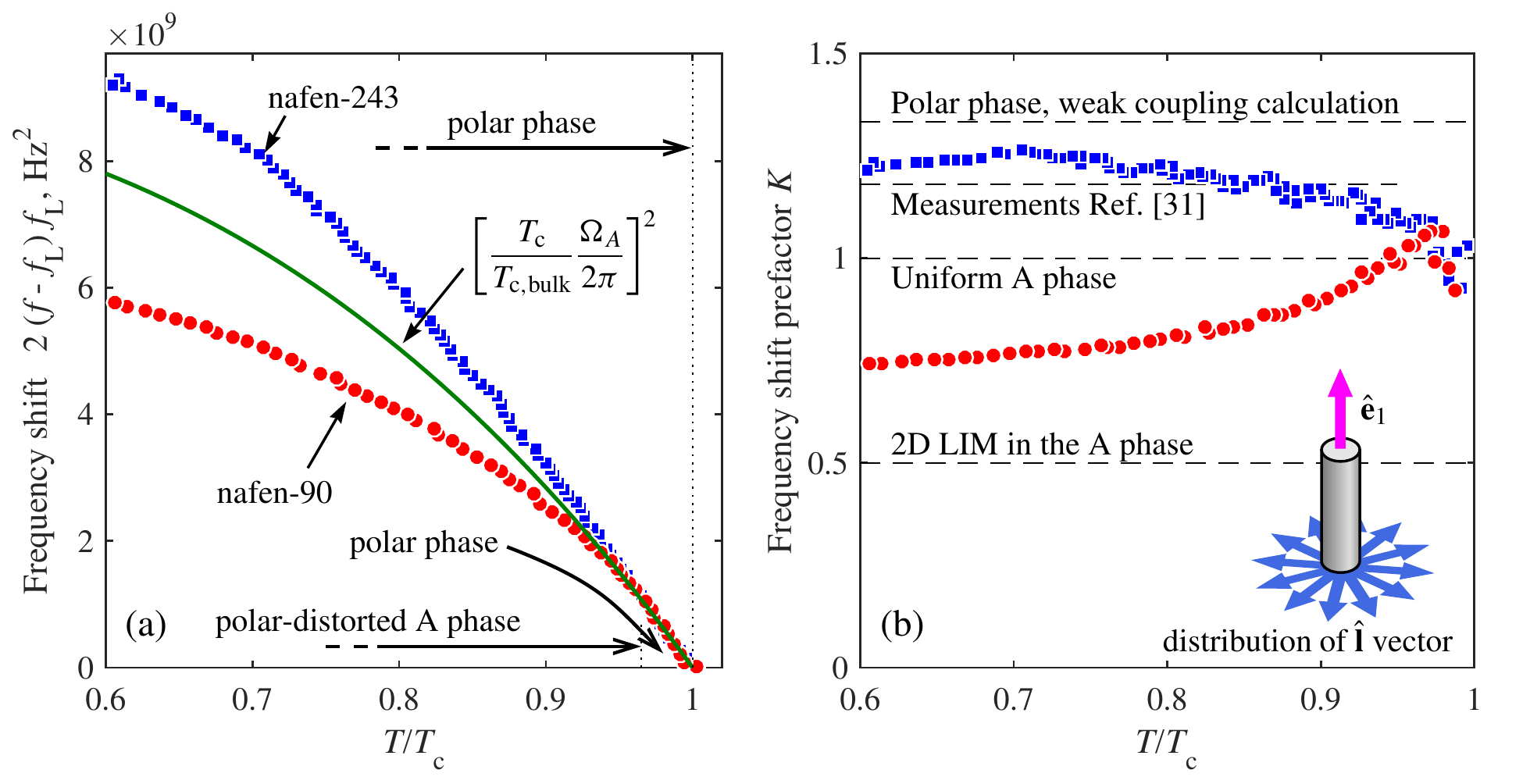}}
\caption{Frequency shift of the NMR response of confined $^3$He as indicator of different order-parameter structures. Measurements are performed at $P = 23.1\,$bar in the magnetic field $\mathbf{H}$ oriented along the nafen strands ($\mu=0$). The corresponding Larmor frequency  $f_{\rm L} = |\gamma| H /2\pi$ is 841\,kHz in nafen-90 and 363\,kHz in nafen-243. In these conditions the NMR spectrum of $^3$He displays a single absorption line at a frequency $f$ shifted from the Larmor frequency $f_{\rm L}$ as $2(f - f_{\rm L}) f_{\rm L} = (\Omega/2\pi)^2$, where $\Omega(T,P)$ is the appropriate Leggett frequency. (a)~Frequency shift as a function of temperature. Temperature is given in units of the normal-superfluid transition temperature of confined $^3$He, which is $T_{\rm c} = 0.982T_{\rm cb}$ for nafen-90 and $T_{\rm c} = 0.97T_{\rm cb}$ for nafen-243 and $T_{\rm cb}$ is the transition temperature in bulk $^3$He. In nafen-243 only the polar phase is seen below $T_{\rm c}$ down to the lowest temperatures. In 
nafen-90 the frequency shift coincides with that in the denser sample in the range $(0.965\div1)T_{\rm c}$ indicating the polar phase, while at lower temperatures it deviates downwards signifying the second-order transition to the polar-distorted A (PdA) phase. At even lower temperature of $0.55T_{\rm c}$ the frequency shifts jumps upwards at the first-order transition to the polar-distorted B phase. The solid line corresponds to $\Omega = \sqrt{K} (T_{\rm c}/T_{\rm cb}) \Omega_A (T\,T_{\rm cb}/T_{\rm c},P)$ with $K = 1$. Here $\Omega_A$ is the Leggett frequency in bulk (undistorted) A phase. (b) The factor $K$ extracted from data in panel (a) compared to earlier measurements and theoretical models. In nafen-243 our measurements agree with the results from Ref.~\onlinecite{Dmitriev2015}, which are slightly below the weak-coupling value $K=4/3$ for the polar phase. In nafen-90 in the PdA phase the value of $K$ drops significantly below 1, expected for the uniform A phase. This signifies formation of the orbital 
LIM glass with $\hat{\mathbf l}$ vectors distributed in the plane transverse to nafen strands. For completely uniform distribution of $\hat{\mathbf l}$ directions one expects $K=1/2$. This is not achieved in our case, probably indicating residual anisotropy of the confinement. }
\label{fig:freqshift}
\end{figure}

As distinct from the $^3$He-A, the polar phase, which appears in the nematically ordered aerogel or nafen \cite{Dmitriev2015}  is not chiral and is time-reversal invariant.
\begin{equation}
A_{\mu i} =\Delta_{\rm P} e^{i\Phi} \hat{d}_{\mu}{\hat e}_i
\,.
 \label{eq:OPHeP}
\end{equation}
The orbital vector $\hat{\bf e}$ is oriented along the strands of nafen. When the disorder is in the form of randomly distributed columnar defects in Fig. \ref{fig:exper}, the polar phase satisfies the Anderson theorem 
\cite{Fomin2018}.  In the same way as the $s$-wave superconductors are insensitive to weak nonmagnetic impurities \cite{Anderson1959}, in the absence of the paramagnetic solid $^3$He the transition temperature to the polar phase  in nafen is almost the same  as $T_{cb}$ in bulk \cite{he4nafen}, see Fig.\ref{fig:freqshift}. It is the only spin-triplet state which has this property. The other phases have lower transition temperatures. On cooling of the polar phase one obtains the polar-distorted A phase --  the chiral state with
\begin{equation}
A_{\mu i} =\Delta_{\rm PdA} e^{i\Phi} \hat{d}_{\mu}(\hat{\bf e}_1 + i b\hat{\bf e}_2)_i
\,,
 \label{eq:OPHeAP}
\end{equation}
where $|b|<1$. 

Let us consider $^3$He-A in Eq.~(\ref{eq:OPHeA}) in isotropic aerogel.
 The superfluid velocity ${\bf v}^{\rm S}$ is determined both by the gradient of the phase $\Phi$ and by the twist of the tetrad field:
\begin{equation}
v_i^{\rm S} = \frac{\hbar}{2m} D_i\Phi = \frac{\hbar}{2m} \left(\nabla_i\Phi  -  \hat{\bf e}_2  \cdot  \nabla_i \hat{\bf e}_1 \right)
\,.
\label{v}
\end{equation}
As follows from Eq.~(\ref{v}) the second term in the rhs plays the role of the vector potential of the synthetic $U(1)$ gauge field in Eq.~(\ref{GaugeField}):
\begin{equation}
A_i  = \hat{\bf e}_2  \cdot  \nabla_i \hat{\bf e}_1
\,,
\label{TetradGaugeField}
\end{equation}
and Eq.~(\ref{GaugeField}) is equivalent to the Mermin-Ho relation \cite{MerminHo1976},
\begin{equation}
(\nabla\times{\bf v}^{\rm S} )^i= \frac{\hbar}{4m} e^{ijk} \hat{\bf l} \cdot \left(    \partial_j \hat{\bf l}  \times    \partial_k \hat{\bf l}\right)
\,.
\label{rotv}
\end{equation} 
In $^3$He-A, several several types of skyrmionic glass state exist: 

(i) First is the equilibrium orbital  LIM glass state  of the orbital vector $\hat{\bf l}$ in Sec.~\ref{OrbGlassSec}. This equilibrium LIM state is obtained by slow cooling from the equilibrium normal (paramagnetic) state through the superfluid transition temperature $T_c$ \cite{Dmitriev2010}.

 (ii) Due to a weak  spin-orbit interaction, the obtained random orientation of the orbital vector $\hat{\bf l}$ in turn serves as the quenched random anisotropy disorder for the spin-nematic vector $\hat{\bf d}$. As a result the equilibrium spin-nematic LIM glass state is formed, with much larger length scale,
$\xi_{{\rm LIM}d} \gg  \xi_D \gg \xi_{{\rm LIM}l}$, where $\xi_D$ is the characteristic length scale of spin-orbit interaction, see Sec.~\ref{CascadeSec}.

 (iii)  There is also the nonequilibrium spin-nematic skyrmion glass state. It is obtained when the large enough resonant continuous radio-frequency
excitation is applied during the cooling through $T_c$ \cite{Dmitriev2010}. The characteristic length scale of this $\hat{\bf d}$ glass is smaller than $\xi_D$, see Sec. \ref{SkyrmionSG}.  In theory such metastable skyrmion glass is obtained by relaxation from the random initial configurations of the order parameter \cite{Chudnovsky2014}.

\section{Orbital LIM glass}
\label{OrbGlassSec}

\subsection{Larkin-Imry-Ma orbital glass in isotropic aerogel}
\label{3DOrbGlassSec}

The LIM state has been realized in the 
chiral Weyl superfluid $^3$He-A immersed in aerogel \cite{Volovik1996,Volovik2008,Dmitriev2010,Halperin2013}.  The random anisotropy of the aerogel strands destroys the long-range orientational order of the  orbital vector $\hat{\bf l}$ giving rise to the LIM disordered state.
In equilibrium LIM state, the singular topological defects, such as singular vortices and the hedgehogs  (analogs of Nambu monopoles \cite{Nambu1977},  which are the end points of the singular Dirac strings \cite{Blaha1976,Volovik1976,Volovik2003}), are absent. Thats is why the LIM state can be characterized by two types of skyrmionic topological charge: the homotopy group   $\pi_2(S^2)=Z$ describing linear skyrmions, and  the homotopy group $\pi_3(S^2)=Z$, which describes the hopfions. According to the Mermin-Ho relation, the $\pi_3(S^2)=Z$ Hopf invariant in Eq.~(\ref{TopInv3}) is expressed in terms of the superfluid helicity \cite{VolovikMineev1977,Ruutu1994,Makhlin1995}:
 \begin{equation}
Q_3=N_{\rm Hopf}=\left(\frac{m_3}{2\pi \hbar} \right)^2 \int d^3x \,{\bf v}^{\rm S}\cdot (\nabla\times{\bf v}^{\rm S})
\,,
\label{N}
\end{equation} 
According to Eq.~(\ref{Q_3^2a}), the typical value of the Hopf invariant in the sample of volume $V$ is $|N_{\rm Hopf}| \sim (V/\xi_{\rm LIM}^3)^{1/2}$, where $\xi_{\rm LIM}$ is the Larkin-Imry-Ma length scale.

\subsection{LIM glass in anisotropic aerogel}
\label{2DOrbGlassSec}

In infinitely stretch aerogel, like nafen which we have in our experiments, the 3D LIM state discussed in Sec. \ref{3DOrbGlassSec} is not realized. Instead, fluctuations of the interstrand distance lead to disordered orientation of the $\hat{\mathbf l}$ vector in the plane perpendicular to the strands. We have indication of such 2D LIM state from the measurements of the Leggett frequency in the polar-distorted A phase confined in nafen, Fig.~\ref{fig:freqshift}. This is the  glass state with the fluctuations of the $Q_2$ charge describing columnar skyrmions in Fig. \ref{objects}. But in the nonequilibrium state it can also contain merons -- the $2\pi$-vortices in the vector $\hat{\bf l}$ field (disclinations), whose winding number is given by
 \begin{equation}
Q_1=\frac{1}{2\pi} \oint ds \, \hat{\bf z}\cdot (\hat{\bf l} \times \partial_s \hat{\bf l}) \,.
\label{Q1l}
\end{equation}  
The core of the vortices with $Q_1=\pm 1$ is soft and is characterized by the half integer topological charge $Q_2=\pm 1/2$. The integral in Eq.~(\ref{Q1l}) is around the soft core.
In bulk $^3$He-A, merons are known as Mermin-Ho vortices, see review \cite{Salomaa1987}.

\subsection{Superfluidity of 3D skyrmion glass in $^3$He-A}
\label{SuperfluidityLIM}

There were several suggestions that in chiral superfluids, superfluidity  can be destroyed by skyrmions \cite{Volovik1996,Coleman2017}.

Let us first consider superfluidity of the LIM state, which has been challenged in Ref.~\onlinecite{Volovik1996}.
According to Eq.~(\ref{A^2}), the average square of superfluid velocity is:
\begin{equation}
\langle{\bf v}_{\rm S}^2\rangle\sim \frac{\hbar^2}{m_3^2\xi_{\rm LIM}^2}
\,,
\label{vs2}
\end{equation}
and thus for $\langle Q_2^2\rangle$  the $m=1$ scaling law in the Eq. (\ref{q_23D})  is applicable. 
The power law $m=1$ in Eq.~(\ref{q_23D}) produces the following scaling for the loop function:
\begin{equation}
\left\langle e^{i \oint_C {\bf A} \cdot d{\bf r} }\right\rangle
\propto 
e^{-L/\xi_{\rm LIM}}
 \,\,,\,\,
L \gg \xi_{\rm LIM}
\,.
\label{LoopFunction}
\end{equation}
According to Ref.~\onlinecite{Toulouse1979}, the linear in $L$ behavior of the exponent of the  loop function means that superfluidity is not destroyed by the LIM texture. The nonzero superfluid density of the LIM state in $^3$He-A has been measured \cite{Nazaretski2004,Bradley2007,Parpia2016}. That is why the LIM state in $^3$He-A represents the system where the off-diagonal long-range order is destroyed. This is a 3D analog of the  2D Berezinskii--Kosterlitz--Thouless superfluid state.  $^3$He-A represents the {\it amorphous topological superfluid}. 
The nonzero superfluid density $\rho_s$ means that the coarse grained $U(1)$ gauge field has a mass. Such glass state with nonzero mass of the effective gauge field corresponds to the confined phase suggested in Ref.~\onlinecite{Vinokur2017}. 

The statement in Ref.~\onlinecite{Volovik1996} has been based on assumption of the $m=2$ scaling law, which is not correct. In case of LIM state, the superfluidity is preserved due to $m=1$ scaling.  But the LIM state can be considered as a heat-insulator phase, since the lowest-energy fermionic states, which live near the Weyl nodes, can be localized. 
In principle,  one may construct (possibly nonequilibrium) states with orbital disorder, in which the mass (charge) superfluidity is lost. Other states are possible when the mass superfluidity is lost, but the spin superfluidity retains, or vice versa:  the spin superfluidity is lost, but the mass superfluidity is not, 
see Sec. \ref{SpinVortexSG}. Such states would provide an analog of the separate charge and spin  localization under random field \cite{Lemut2017}.  However, it is not excluded that whatever is the scaling law, the glass state remains superfluid because of the pinning of the texture. 

\subsection{Skyrme superfluid vs Skyrme insulator}

Another theoretical challenge is the stability of superflow in pure $^3$He-A. It has been suggested that easy creation of skyrmions by the mass current destroys the superfluidity, and possible corresponding  nonsuperfluid state has been called {\it Skyrme insulator} \cite{Coleman2017}.
In reality, however, a finite-size system remains superfluid since the Feynman critical velocity $v_{\rm Feynman}$, approximately inversely proportional to the system size, is not zero. In a channel of finite thickness, both the creation of skyrmion in $^3$He-A and creation of vortex ring in superfluid $^4$He require the overcoming of the critical velocity, at which the creation of these objects become energetically favorable.  The Feyman critical velocity for creation of a vortex ring in superfluid $^4$He is
$v_{\rm Feynman} \sim (\hbar/ md) \ln(d/a)$, where  $d$ is the width of the channel or slab, and $a$ is the core size of the vortex.  
For skyrmions, the core size $a\sim d$, and $v_{\rm Feynman} \sim \hbar/ md$, which is only logarithmically smaller than in superfluid $^4$He.
 
The instability of the supercurrent towards creation of skyrmions has been measured in $^3$He-A, see discussion in Ref.~\onlinecite{Volovik2003}. The measured threshold is much larger than the Feyman critical velocity. The reason is that while the creation of the skyrmions is energetically becomes favorable, the superflow is locally stable and the potential barrier for creation is by many orders of magnitude larger than the temperature of the system. That is why the skyrmions are created at the critical velocity, at which the helical instability of the orbital texture develops.  In principle, one can construct  geometry in which the superflow is locally unstable. In this case the critical velocity will be reduced to the Feynman critical velocity.

\section{Spin glasses}
\label{SpinGlassSec}

\subsection{Cascade of LIM processes}
\label{CascadeSec}

The hierarchy of energy scales and corresponding length scales produces the {\it cascade LIM processes}: the quenched orientational disorder of aerogel strands on nanoscales gives rise to the orientational disorder in the orbital vector field ({\it orbital glass state}) on a microscale, which in turn leads to the spin disorder  ({\it spin glass state}) on a milliscale.
According to the NMR measurements \cite{Dmitriev2010}, the LIM scale for the disorder of the orbital vector $\hat{\bf l}$ is smaller than the characteristic scale of spin-orbit interaction, $\xi_{LIM\hat{\bf l}} \ll \xi_D$. Then the corresponding LIM scale of the disordered state of the spin-nematics vector $\hat{\bf d}$ is:
 \begin{equation}
\xi_{LIM\hat{\bf d}}= \frac{\xi_D^4}{\xi_{LIM\hat{\bf l}}^3} \gg \xi_D \gg \xi_{LIM\hat{\bf l}} 
\,.
\label{xiLIMd}
\end{equation} 

This equilibrium {\it spin glass state} can be characterized by its own  $\pi_2(S^2)$ and $\pi_3(S^2)$ topological numbers. The latter is the spin Hopf invariant:
 \begin{equation}
N_{\rm Hopf}= \frac{1}{32\pi^2} \int d^3x \,e^{ijk} A_i F_{jk}
\,,
\label{Nspin}
\end{equation} 
where 
 \begin{equation}
 F_{jk}= \partial_j A_k - \partial_k A_j=  \hat{\bf d} \cdot \left(\partial_j \hat{\bf d}  \times    \partial_k \hat{\bf d}\right)
\,.
\label{Bd}
\end{equation} 
So  this combination  of the  {\it orbital glass} and  {\it spin glass} represents the {\it hierarchical double Skyrme glass}.

\begin{figure}[t]
\centerline{\includegraphics[width=0.7\linewidth]{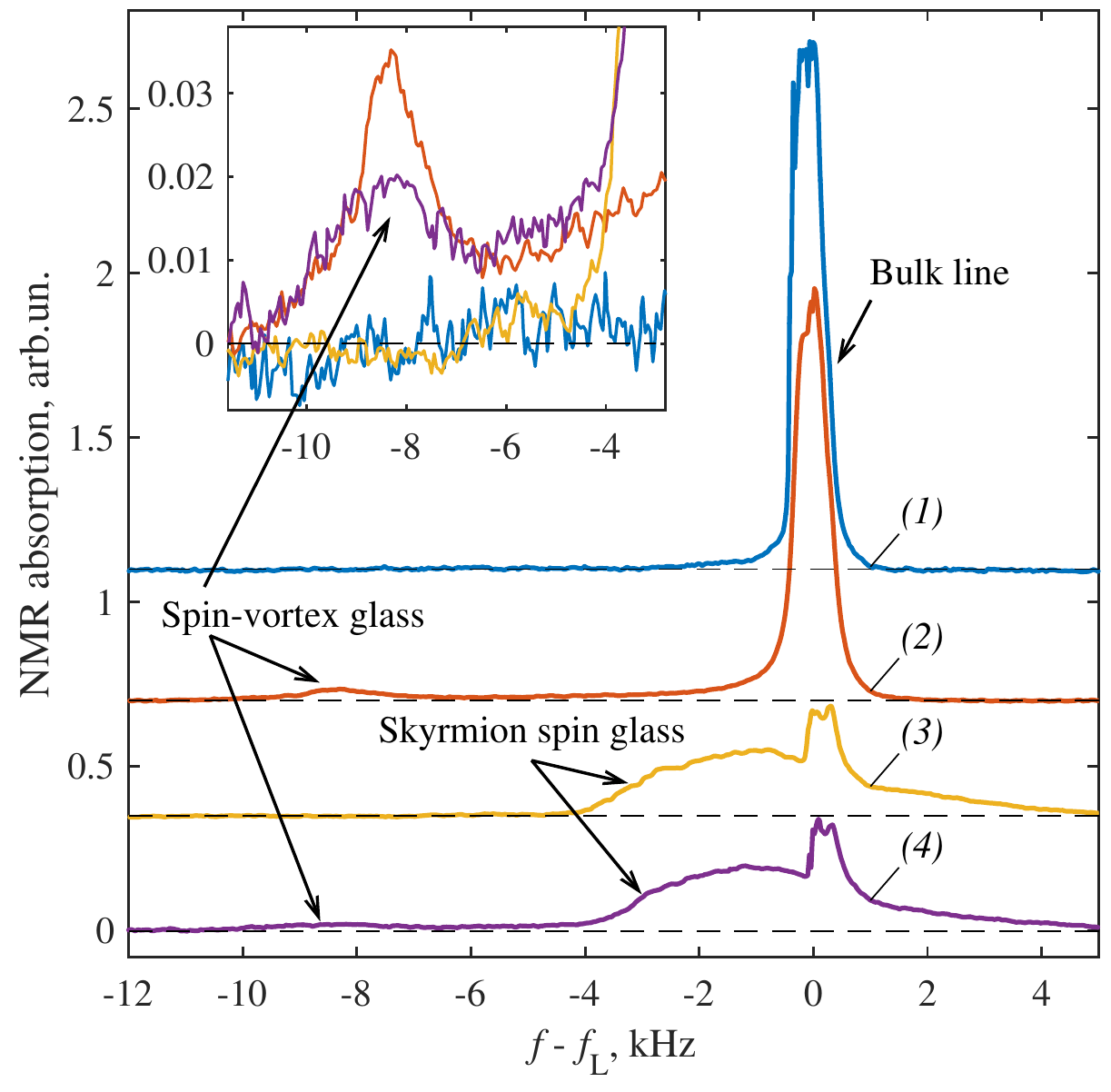}}
\caption{Spin glasses in the polar-distorted A (PdA) phase of superfluid $^3$He in NMR observations. The main panel shows NMR absorption spectra measured at temperature $T = 0.4T_{\rm c}$ and pressure $P = 29.5\,$bar for $^3$He confined in nafen-90 with the magnetic field $\mathbf{H}$ transverse to nafen strands ($\mu = 90^\circ$). The horizontal axis is the shift of the frequency $f$ of the NMR response from the Larmor frequency $f_{\rm L}  = 409\,$kHz. On cooling from the normal phase  at this pressure and confinement, first the transition to the polar phase occurs, which is followed by the second-order transition to the PdA phase, similar to Fig.~\ref{fig:freqshift}. \textit{(1)}  The spectrum is measured after normal to superfluid transition happened in the transverse magnetic field ($\mu=90^\circ$) and no rf pumping. In this case disordered spin structures are not created. The spectrum includes only the bulk line at zero frequency shift. \textit{(2)} Here strong rf drive is applied during the superfluid 
transition in the field tilted at $\mu=20^\circ$ and the spin-vortex glass is formed, while formation of the HQV glass is suppressed. The spin-vortex glass is manifested by a satellite peak at $f - f_{\rm L} \approx - \Omega_{\rm PdA}^2/4\pi^2 f_{\rm L} \approx -8\,$kHz, where $\Omega_{\rm PdA}$ is the Leggett frequency in the PdA phase. The satellite peak originates from spin $\hat{\mathbf{d}}$ solitons stretched between spin vortices. The solitons have thickness of the dipolar length $\xi_D\sim 10\,\mu$m and occupy relatively small part of the sample volume. \textit{(3)} Skyrmion spin glass is formed when the state \textit{(1)} is cooled further through the first-order phase transition to the polar-distorted B (PdB) phase and then warmed through the first-order phase transition from the PdB to PdA phase. In contrast to spin-vortex glass in \textit{(2)}, the satellite appears at less negative frequency shifts and is also wider and larger in intensity, since essentially the whole volume contributes to 
absorption. \textit{(4)} When the spin-vortex glass in \textit{(2)} is cycled to the PdB phase and back, the combined spin-vortex and spin-skyrmion glasses are created, as seen from the two satellites present in the spectrum. Note that the $\hat{\mathbf{d}}$-soliton satellite is modified when the soliton is embedded in the skyrmion spin glass. \textit{Inset} shows zoomed view of the spectral region with the $\hat{\mathbf{d}}$-soliton satellite. }
\label{fig:spectra}
\end{figure}

\subsection{Skyrmion spin glass}
\label{SkyrmionSG}

The nonequilibrium skyrmion glass state originally has been obtained when the large enough resonant continuous radio-frequency excitation has been applied during the cooling through $T_c$ from the normal state to $^3$He-A \cite{Dmitriev2010}.  The NMR signature of this state demonstrates that the characteristic length scale of textures in this $\hat{\bf d}$-glass is smaller than $\xi_D$, contrary to the equilibrium spin glass in Eq.~(\ref{xiLIMd}).  The nonequilibrium skyrmion glass with the same NMR signature can be also obtained by warming from the B phase to the A phase through the first-order phase transition, see Fig.~\ref{Table} and spectra \textit{(3)} and \textit{(4)} in Fig.~\ref{fig:spectra}. Since such spin glass 
exists due to spin-orbit interaction with orbital spin glass, which disappears on transition from the A phase to the polar phase, also the spin-skyrmion glass is annealed on this transition. On return back from the polar phase to the A phase we observe change from spectra \textit{(3)} and \textit{(4)} to \textit{(1)} and \textit{(2)}, respectively.

\subsection{Spin-vortex glass and spin-current confinement}
\label{SpinVortexSG}

Spin vortices in the $^3$He-A and in the polar phase are vortices in the $\hat{\bf d}$-field in the presence of large enough magnetic field. If the characteristic magnetic length $\xi_{\rm magn} \ll \xi_D$, the magnetic field orients $\hat{\bf d}\perp {\bf H}$, and the spin vortices are described by $\pi_1(S^1)=Z$ winding number as in Eq.~(\ref{Q1l}):
 \begin{equation}
Q_1=\frac{1}{2\pi} \oint ds \, \hat{\bf z}\cdot (\hat{\bf d} \times \partial_s \hat{\bf d}) \,.
\label{Q1}
\end{equation}  
Here the integral is  around the vortex line.
The core size of spin vortex is determined by magnetic length, $\xi_{\rm magn}$.
Such a smooth core of spin vortex represents meron in the $\hat{\bf d}$-field, the half of the skyrmion described by the half-integer topological charge:
 \begin{equation}
Q_2=\frac{1}{4\pi} \int dx\,dy\,  \hat{\bf d} \cdot \left(\partial_x \hat{\bf d}  \times    \partial_y \hat{\bf d}\right)
=\pm 1/2 \,.
\label{Merons}
\end{equation}

Our NMR experiments suggest that spin-vortex glass is formed after phase transition from the normal state to the polar phase when strong pumping (sufficient to significantly saturate normal-state response) is applied during the transition.  This spin-vortex glass is preserved on the transition from the polar phase to the A phase (where spin vortices are probably  pinned by the orbital LIM texture due to spin-orbit interaction) and back, see Fig.~\ref{Table}. In experiment spin-vortex glass is seen via characteristic satellite in the NMR spectra at a relatively large negative frequency shift, see spectra \textit{(2)} and \textit{(4)} in Fig.~\ref{fig:spectra}.

Let us consider fluctuations of the topological charge in the spin-vortex state.
We introduce the effective gauge field $\mathbf{A}$ describing the $U(1)$ spin vortices with density of topological charge $q_1$. This is similar to the effective gauge field representing the equivalent description of disorder in terms the distributed linear topological defects in e.g. Refs.~\onlinecite{DzyaloshinskiiVolovik1978,DzyaloshinskiiVolovik1980}, where in particular the spin glass  has been  treated in terms of the effective $SU(2)$ gauge field. The noise in the distribution of positively charged $Q_1=+1$ and negatively charged $Q_1=-1$ spin vortices gives:
\begin{equation}
 \langle Q_1^2\rangle= \left(\oint {\bf A }\cdot d{\bf x}\right)^2 =  \left(\int dS \, q_1\right)^2  \sim L^2/\xi_{\rm LIM}^2
\,.
\label{q_13D}
\end{equation}
Now the power law is $m=2$, and Eq.~(\ref{q_13D}) gives the following scaling for the loop function:
\begin{equation}
\left\langle e^{i \oint_C {\bf A} \cdot d{\bf r} }\right\rangle
\propto 
e^{-L^2/\xi_{\rm LIM}^2}
 \,\,,\,\,
L \gg \xi_{\rm LIM}
\,.
\label{LoopFunction2}
\end{equation}
Such behavior suggests that  in spin-vortex glass the spin superfluidity is destroyed, as distinct from the spin skyrmion state with $m=1$.
In gauge theories, the state with the area law is the confinement phase, because the corresponding charges are confined there,
see e.g. the book by Polyakov \cite{Polyakov1987}. Again, the role of the pinning remains unclear: the strong pinning of the topological defects may or may not restore the spin superfluidity.

\section{Vortex glasses}
\label{VortexGlassSec}

\subsection{Glass of half-quantum vortices}

\begin{figure}
\centerline{\includegraphics[width=0.75\linewidth]{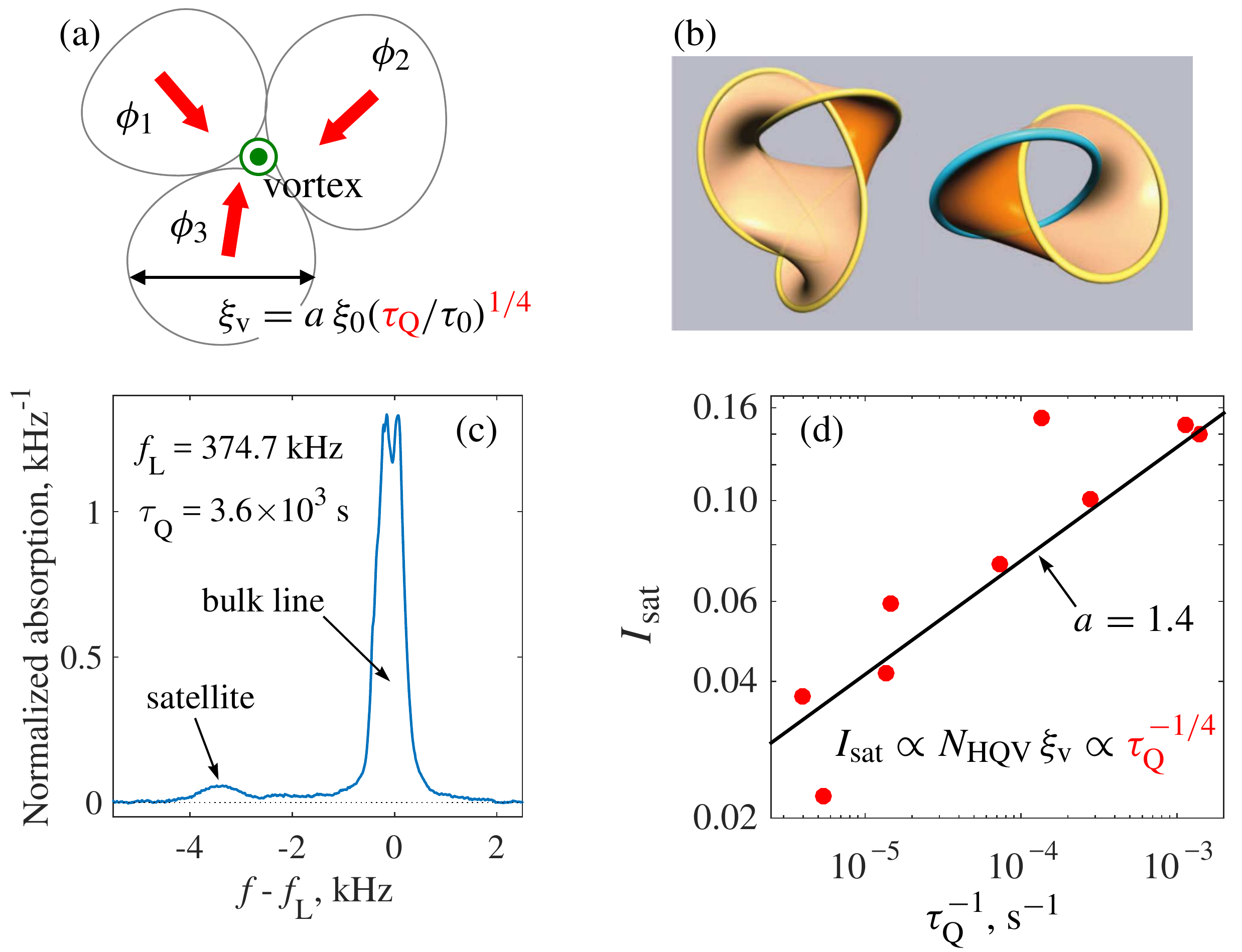}}
\caption{Glass of half-quantum vortices created by the Kibble-Zurek mechanism in the polar phase of superfluid $^3$He. (a) After non-equilibrium phase transition with a finite rate $\tau_Q$ of the temperature sweep through $T_{\rm c}$, the phase $\phi$ of the superfluid order parameter can develop to different values in the casually disconnected regions. When such regions meet, a vortex can be trapped if the phase winding is $2\pi$ for conventional superfluids or $\pi$ for HQVs in the polar phase. In unconfined superfluid, KZ vortices rapidly decay \cite{Bunkov2014}, but in the polar phase they are pinned on the nafen strands and remain at their initial density determined by the KZ length $\xi_{\rm v}$. (b) In a magnetic field, transverse to the strands, $\hat{\mathbf{d}}$ solitons emerge between vortex segments with opposite orientation of spin-current circulation. For disordered and interlinked loops these solitons form Seifert surfaces. The examples here are from Ref.~\onlinecite{vanWijk2006}. (c) Spin 
waves bound to $\hat{\mathbf{d}}$ solitons give rise to a characteristic satellite in the NMR spectrum. The measurement here is done at $T=0.69 T_{\rm c}$ and $P=7\,$bar. (d) The normalized area of the satellite $I_{\rm sat}$ is proportional to the number of HQVs $N_{\rm HQV} \propto \xi_{\rm v}^{-2}$ and to the average soliton length $\xi_{\rm v}$. The experimental points (circles) \cite{Autti2016a} follow the expected dependence $I_{\rm sat} \propto \tau_Q^{-1/4}$. Moreover, they are in a reasonable agreement with the theoretical expectation (solid line) based on the value of $a$ measured in $^3$He-B \cite{Bauerle1998} and no further fitting parameters. }
\label{fig:kz}
\end{figure}

Quantized vortices strongly pinned by nafen strands form the vortex glass. The vortex glass is obtained with the Kibble-Zurek (KZ) mechanism by fast cooling through $T_c$ to the polar phase, Fig.~\ref{fig:kz}(a). As follows from the NMR experiments, the vortex glass consists of the pinned Alice strings -- half-quantum vortices (HQVs) with the following structure of the order parameter:
\begin{equation}
A_{\mu i} =\Delta_P e^{i\varphi/2}\left(\hat{x}_{\mu}\cos \frac{\varphi}{2} +\hat{y}_{\mu}\sin \frac{\varphi}{2}\right) {\hat e}_i
\,,
 \label{eq:OPhalf}
\end{equation} 
where $\varphi$ is the azimuthal angle around the vortex line. The half-quantum vortex is a combination of the mass vortex with winding number $N=1/2$ and spin half-vortex with $Q_1=1/2$. That is why, when the magnetic field is switched on, the spin half-vortex gives rise to the spin soliton terminating on the vortex, Fig.~\ref{fig:kz}(b). The spin solitons formed between HQVs produce the satellite peak in NMR spectrum, Fig.~\ref{fig:kz}(c).
The intensity of the peak allows us to estimate the density of  half-quantum vortices, which agrees with the expectations from the KZ mechanism, see Fig.~\ref{fig:kz}(d).

The observed vortex glass -- the {\it Alice glass} -- differs from the Larkin vortex glass in superconductors, where vortices have preferred  orientation of magnetic fluxes along the magnetic field.
In the isotropic aerogel vortices have random  orientation of vortex lines. In the aerogel with preferrable orientations of the strands the vortices form the disordered Ising glass with the random distributions of the 
winding numbers $N=+1$ and $N=-1$. 
This type of vortex matter adds to the Zoo of vortex states in superconductors: Bragg glass, vortex glass, vortex liquid and the Abrikosov 
lattice \cite{Biedenkopf2005,Biedenkopf2007}.

 Several types of the solitonic glass are possible. Spin solitons are formed in the glass or the lattice of half-quantum vortices. They are formed between the half-quantum vortices due to spin-orbit interaction. In the vortex glass they form the solitonic glass, and in the vortex lattice  --  the analog of Bragg glass ({\it solitonic Bragg glass}). 

\section{Discussion. Topological fermionic glasses}
\label{FermGlassSec}

The spin-triplet superfluid phases of $^3$He have rich topological properties, which allow us to produce many types of the glass states classified in terms of the pinned topological defects (Alice strings, monopoles, domain walls, etc.) and textures (skyrmions, hopfions, merons, solitons, etc.). Some of these states have been experimentally identified in NMR experiments, but many other states are still waiting for their strong identification. Experimental and theoretical study  of these states may lead to discovery of new phenomena and new concepts in the physics of the topological disorder. 

However, what seems to be the most important, is that all the observed spin-triplet superfluid phases of $^3$He are topological superfluids, described by the topological invariants in momentum space. 

1) The A phase and the polar-distorted A phase are Weyl superfluids with Weyl nodes in the fermionic spectrum. The Weyl points serve as the Berry phase magnetic monopoles, but now in momentum space \cite{Volovik1987b}. The corresponding Hamiltonian for quasiparticles near the Weyl points has the form:
$H=e^i_a(p_i - qA_i)\sigma^a$, where $\sigma^a$ are the Pauli matrices in the Bogoliubov-Nambu particle-hole space; $e^i_a$ are the elements of the effective (synthetic) tetrad field; ${\bf A}=k_F\hat{\bf l}$ is the effective (synthetic) electromagnetic field; and $q=\pm 1$ is effective electric charge.

2) The polar phase has Dirac nodal line  in the fermionic spectrum and correspondingly the degenerate tetrad field \cite{NissinenVolovik2018}. 

3) The B phase and the polar-distorted B phase are fully gapped topological superfluids of the DIII class with Majorana fermions on the surface. These phases become the higher-order topological superfluids in applied magnetic field, see e.g. Refs. \cite{Volovik2010,Khalaf2018} 

In the glass phases the disorder adds new features to the topological structure in momentum space, and the momentum-space topology meets the real-space topology \cite{Horava2005,Prodan2016,ShouChengZhang2017,DisorderReview2016}.

In particular, in the disordered LIM state of $^3$He-A in isotropic aerogel, the positions $\pm k_F\hat{\bf l}$ of the Weyl nodes and the orientations of the tetrads  $e^i_a$ are smoothly and randomly distributed in space  forming a unique example of a {\it Weyl glass}.
The random positions of the nodes give rise to the random effective gauge field ${\bf A}=k_F\nabla\times \hat{\bf l}$, while the random orientations of the tetrads with $\left\langle e_a^\mu \right\rangle=0$ form the analog of the
{\it torsion foam} in quantum gravity \cite{Hawking1978,HansonRegge1979}.

Smooth disorder of superfluid phases of $^3$He allows us to consider the disordered state as collection of domains with different values of the momentum space invariants -- the Chern numbers
 \cite{Volovik2018,Lian2018}. Such topological glass state represents the real space analog of the Chern mosaic in the space of parameters \cite{Ojanen2016,Ojanen2017}.

{\bf Acknowledgements}. We thank Vladimir Dmitriev, Eugene Chudnovsky, Alexei Yudin and Samuli Autti for useful discussions. This work has been supported by the European Research Council (ERC) under the European Union's Horizon 2020 research and innovation programme (Grant Agreement No. 694248).

\end{document}